\documentclass[aps,prd,twocolumn,groupedaddress,showpacs]{revtex4}
\usepackage{epsf,epsfig,graphics,graphicx}
\begin{document}

\title{Decaying Superheavy Dark Matter and Subgalactic Structure
of the Universe}

\author{Chung-Hsien Chou}
\email{chouch@phys.sinica.edu.tw}
\author{Kin-Wang Ng}
\email{nkw@phys.sinica.edu.tw}
\affiliation{Institute of Physics, Academia Sinica, Nankang, Taipei,
Taiwan 11529, R.O.C.}

\date{\today}

\begin{abstract}
The collisionless cold dark matter (CCDM) model predicts overly dense cores
in dark matter halos and overly abundant subhalos. We show that the idea
that CDM are decaying superheavy particles which produce ultra-high energy
cosmic rays with energies beyond the Greisen-Zatsepin-Kuzmin cutoff may
simultaneously solve the problem of subgalactic structure formation
in CCDM model. In particular, the Kuzmin-Rubakov's decaying superheavy
CDM model may give an explanation to the smallness of the cosmological constant
and a new thought to the CDM experimental search.
\end{abstract}

\pacs{95.35.+d, 98.62.Gq, 98.70.Sa, 98.80.Cq}
\maketitle

\section{Introduction}

Recent cosmological observations such as dynamical mass, Type Ia
supernovae, gravitational lensing, and cosmic microwave background
anisotropies, concordantly predict a spatially flat universe
containing a mixture of $5\%$ baryons, $25\%$ cold dark matter
(CDM), and $70\%$ vacuum-like dark energy~\cite{wang,wmap}, termed
as the standard $\Lambda$CDM model. The identities and the nature
of dark matter and dark energy are among some of the biggest
puzzles in contemporary physics.

Although the nature of CDM is yet unknown, it is successfully
treated in many aspects as weakly interacting particles. However,
there exist serious discrepancies between observations and
numerical simulations of CDM halos in collisionless cold dark
matter (CCDM) models~\cite{nfw,moore,ccdm}, which predict too much
power on small scales, manifested as cuspy CDM cores in dwarf
galaxies~\cite{dwarf}, galaxies like the Milky Way~\cite{galaxy},
and central regions of galaxy clusters~\cite{cluster} as well as a
large excess of CDM subhalos or dwarf galaxies within the Local
Group~\cite{ccdm}.

To alleviate the discrepancies, among many other attempts, models of
non-standard interacting CDM have been proposed. They include
self-interactions~\cite{sper}, annihilations~\cite{kap},
and decaying cold dark matter (DCDM)~\cite{cen,sal}.
Although these models involve different interactions, almost all interactions
result in an adiabatic expansion of the cuspy halo that lowers the core density
and reduces the number of subhalos.
However, both self-interacting and annihilating CDM models require
embarrassing large interaction cross-sections that have made the models
less appealing. Although DCDM models are viable, possible underlying
particle physics has been ignored.

Another big puzzle in astrophysics is the origin of the ultra-high
energy cosmic rays (UHECR).  One may expect that UHECR should
originate from some unknown astrophysical sources at extragalactic
scales. Greisen, Zatsepin, and Kuzmin (GZK)~\cite{gzk} observed
that due to inverse Compton scatterings of the relic photons the
UHECR energy spectrum produced at cosmological distances should
steepen abruptly  at energy $\sim 10^{10}$ GeV. However, a number
of cosmic ray events with energies beyond the GZK cutoff have been
observed by Fly's Eye~\cite{fly} and AGASA~\cite{agasa}. A simple
solution to this impasse is to invoke new physics in which UHECR
can be produced in a cosmologically local part of the Universe.
Ideas such as long-lived metastable superheavy particles that are
decaying at the present epoch~\cite{ellis,bir,ber,kuz,fod},
annihilations of stable supermassive particles in
halos~\cite{bla}, and collapses of cosmic topological
defects~\cite{bha} have been proposed. In most of the models the
superheavy objects can simultaneously be viable candidates for DM.

In this paper, we try to address these issues at the same time
within a single theoretical framework. We pursue the DCDM
scenario, suggesting that the CDM is decaying weakly interacting
superheavy particles with mass of the grand unification scale. In
our scenario, not only the decay would produce much less
concentrated cores in CDM halos, but also the decay products
contain highly energetic quarks and leptons which lead to the
production of ultra-high energy cosmic rays (UHECR) with energies
beyond the Greisen-Zatsepin-Kuzmin cutoff. Moreover, the longevity
of the superheavy particles may shed new light on the origin of
the observed small value of the cosmological constant.

The paper is organized as follows: In section \textbf{II} we
illustrate our idea by using the Kuzmin-Rubakov model. After
briefly reviewing this model, we show in section \textbf{III} how
this model can be naturally fitted into the scenario of DCDM. We
show how this model solves the cuspy halo problem, and find out
the parameter space which allow us to solve the origin of UHECR as
well. In section \textbf{IV} we discuss some phenomenological
implications and suggest that some on-going experiments could test
this scenario.

\section{Kuzmin-Rubakov model}

Here we will concentrate on a specific scenario proposed by Kuzmin
and Rubakov (KR)~\cite{kuz} and show how the KR scenario for
producing UHECR is related to the subgalactic structure of the
Universe.

KR~\cite{kuz} considered an extended standard model with a new $SU(2)_X$
gauge interaction and two left-handed $SU(2)_X$ fermionic
doublets ${\bf X}$ and ${\bf Y}$ and four right-handed singlets.
Here at least two doublets are introduced because the $SU(2)_X$ anomaly
prevents the number of $SU(2)_X$ doublets from being odd.
All new particles are singlets of the standard model, while some conventional
quarks and leptons may carry non-trivial $SU(2)_X$ quantum numbers.
The $SU(2)_X$ gauge symmetry is assumed to be broken at certain high energy
scale, giving large masses $m_{X,Y}$ to all ${\bf X}$ and ${\bf Y}$ particles.
Furthermore, ${\bf X}$ and ${\bf Y}$ are assumed to carry different
global symmetries, so there is no mixing between them. As such, both the
lightest of ${\bf X}$ and the lightest of ${\bf Y}$, which we call $X$ and $Y$
respectively, are perturbatively stable. However, $SU(2)_X$ instantons induce
effective interactions violating global symmetries of $X$ and $Y$.
Assume $m_X > m_Y$, then the instanton effects lead to the decay
\begin{equation}
X \rightarrow Y + {\rm quarks} + {\rm leptons}
\label{decay}
\end{equation}
with a long lifetime roughly estimated as $\tau_X \sim m_X^{-1}
e^{4\pi/\alpha_X}$, where $\alpha_X$ is the $SU(2)_X$ gauge
coupling constant. With the choices $m_X \agt 10^{13} {\rm GeV}$
and $\alpha_X \alt 0.1$, $\tau_X \agt 10 {\rm Gyrs}$ and $X$
particles are decaying at the present epoch. There have been many
discussions on the production of $X$ particles in the early
Universe. $X$ particles may be produced thermally during reheating
after inflation with the produced energy density comparable to the
critical energy density of the Universe~\cite{kuz} (see also
Refs.~\cite{ber,chu}). Also, it was realized in the same or
different context that superheavy particles can be efficiently
generated from vacuum quantum fluctuations during
inflation~\cite{enq} or couplings to the inflaton field during
preheating~\cite{baa}.

The particles $X$ and $Y$ are good dark matter candidates.
According to KR, there are two possible outcomes after $X$
particles have decayed. If $Y$ particles are perturbatively
stable, they are also stable against instanton-induced
interactions in virtue of energy conservation and instanton
selection rules. In addition, if $m_X \agt m_Y$, an approximately
equal amount of $Y$ particles is produced in the early Universe.
Therefore, the decay products would contain stable supermassive
$Y$ particles that constitute a dominant fraction of the CDM with
a small admixture of $X$ particles as well as highly energetic
quark jets and leptons that subsequently produce UHECR.
Alternatively, the Higgs sector and its interactions with fermions
may be organized in such a way that Y particles are in fact
perturbatively unstable. As such, $Y$ particles would instantly
decay into relativistic particles and leave metastable $X$
particles being the CDM.

Intriguingly, it has been recently pointed out that if the
longevity of the superheavy particles in the KR model is due to
instanton-induced decays, the observed small but finite
cosmological constant can be explained by instantons or vacuum
tunnelling effects in a theory with degenerate vacua~\cite{jai}.
In such a theory, the vacuum energy density of the true ground
state is smaller than that in one of the degenerate vacua where we
live now by an exponentially small amount if quantum tunnelling
between the degenerate vacua is allowed~\cite{yok}.

\section{ Resolution of the cuspy halo problem and UHECR}

We now turn to the cuspy halo problem and show how this problem
can be solved within the context of the KR model. Numerical
simulations of CCDM halos show cuspy halo density profiles well
fit with the generalized Navarro-Frenk-White (NFW)
form~\cite{nfw,moore,ccdm},
\begin{equation}
\rho(r)=\rho_c \left(\frac{r}{r_c}\right)^{-\alpha}
        \left(1+\frac{r}{r_c}\right)^{\alpha-3},
\label{prof}
\end{equation}
with the slope parameter $\alpha\simeq 1-1.5$
and the concentration parameter $c\equiv r_{200}/r_c \simeq 20$,
where $r_c$ is the core radius, $\rho_c$ is the mean density of the Universe
at the time the halo collapsed, and $r_{200}$ is the radius within which
the mean density $\rho_{200}$ is 200 times the present mean density of
the Universe. However, observations indicate flat core density profiles
with $\alpha \alt 0.5$ and smaller concentrations with
$c\simeq 6-8$~\cite{dwarf,galaxy,cluster}. Below we will
simply study the effect of DCDM to the original NFW profile
with $\alpha=1$~\cite{nfw} in Eq.~(\ref{prof}). Defining $x=r/r_{200}$, it
gives the halo mass profile $M(x)=M_{200} F(x)$ that is the mass within $x$
and the associated rotational velocity $V(x)=V_{200}[F(x)/x]^{1\over2}$,
where $M_{200}=M(x=1)$, $V_{200}=V(x=1)$, and
\begin{equation}
F(x)=[\ln(1+cx)-cx/(1+cx)]/[\ln(1+c)-c/(1+c)].
\label{mv}
\end{equation}

Suppose a CDM halo gas composed of $X$ particles is formed at some
high redshift with the NFW profile and a velocity dispersion
$v_X=\sqrt{G M_{200,X}/2r_{200,X}}$, where $M_{200,X}$ is the mass
of X particles within the radius $r_{200,X}$. The observed
velocity dispersion typically ranges from 10 to 1000 km/s for
dwarf halo to cluster halo. In $X$'s rest frame, the
decay~(\ref{decay}) produces a $Y$ with a recoiling velocity
$\gamma_{rc} v_{rc}=\delta(1-\delta/2)/(1-\delta)$, where
$\gamma_{rc}=1/\sqrt{1-v_{rc}^2}$ and $\delta=(m_X-m_Y)/m_X$, and
highly relativistic quarks and leptons of energy
$E_{q,l}=\gamma_{rc} v_{rc} m_X (1-\delta)$. The value of $\delta$
depends on the detail dynamics of the high energy model. Here we
will treat it as an input parameter. There are two possibilities.
When $1 \agt \delta > v_X$, we find that $Y$ would be relativistic
and/or beyond the escape velocity of the halo. This together with
the case of an unstable $Y$ correspond to the scenario discussed
in Ref.~\cite{cen}, to which readers may refer for details. In the
following, we will discuss the case for $\delta < v_X$, i.e.
nearly degenerate masses, in which stable $Y$ particles would be
bound to the halo with an averaged velocity about $\sqrt{
v_X^2+v_{rc}^2}$ ($v_{rc}\simeq \delta$) just after the decay of
$X$ particles. In particular, $\delta\simeq  1-2\times 10^{-4}$
corresponds to the case considered in Ref.~\cite{sal}.

Let us assume that most $X$ particles have decayed and that the halo
of $Y$ particles with the NFW profile has been formed by now.
Using the virial theorem it can be shown that the core radius has expanded to
\begin{equation}
r_{c,Y}=r_{c,X}/y;\quad  y\equiv\frac{1-2\delta}{1-\delta}-
\frac{\delta^2}{v_X^2} \frac{(1-\delta/2)^2}{(1-\delta)^3}.
\label{yhalo}
\end{equation}
We will follow the method in Ref.~\cite{cen} to work out the
consequences of this core expansion. The difference is that here
the mass inside $r_{200,X}/y$ is only slightly changed to
$(1-\delta)M_{200,X}$. As such, the final density within
$r_{200,X}/y$ is $y^3(1-\delta)\rho_{200}$. To obtain $r_{200,Y}$,
we solve for $r=yr_{200,Y}$ in Eq.~(\ref{prof}) ($\alpha=1$)
within which the initial density is
$y^{-3}(1-\delta)^{-1}\rho_{200}$. The resulting equation is $x^3
F^{-1}(x)=y^3(1-\delta)$ and we find that $r_{200,Y}\simeq
y^{0.2}r_{200,X}$ for $y\alt 1$ and $\delta<<1$. Hence we obtain
$c_Y\simeq y^{1.2}c_X$. To circumvent the over-concentration
problem, $y$ should be about 0.4, implying that $\delta\sim
0.77v_X$. Using $c_X=20$, $y=0.4$, and Eq.~(\ref{mv}), we obtain
the mass profiles and rotation curves of the original $X$ halo and
the presently formed $Y$ halo shown in Fig.~1. We find that
$M_{200,Y}\simeq 0.58 M_{200,X}$ and $M_Y(r=0.1r_{200,Y})\simeq
0.27 M_X(r=0.1r_{200,X})$, and that $V_{200,Y}\simeq 0.83
V_{200,X}$, $V_{max,Y}\simeq 0.64 V_{max,X}$, and $r_{max,Y}\simeq
2.5 r_{max,X}$, where $V_{max}$ is the maximum rotational velocity
at radius $r_{max}$. In Fig.~1, we have also reproduced the mass
profile and the rotation curve for the case~\cite{cen} in which
$X$ decays into relativistic particles. This requires solving for
$x=r/r_{200,X}$ in the equation $x^3 F^{-1}(x)=y^4$, where the
mass inside $r_{200,X}/y$ is $yM_{200,X}$ and $y=0.5$ is the
fraction of $X$ particles that still remain by now. In this case,
the softening of the central concentration is the same as in the
$Y$ halo, but the reduction in the halo mass profile and the
flattening of the rotation curve are even more pronounced. Thus we
have shown that one can put KR model which was originally proposed
to explain the origin of UHECR into the DCDM model.

\begin{figure}
\leavevmode
\hbox{
\epsfxsize=3.5in
\epsffile{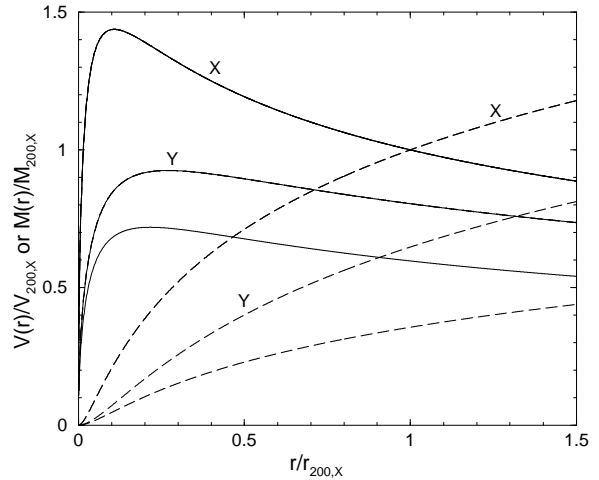}}
\caption{Solid (dashed) curves represent respectively from up to down the
rotation curves (mass profiles) for the $X$ halo which is from the NFW
profile in the CCDM model, the $Y$ halo in the DCDM model,
and the case in which $X$ decays into relativistic particles. The x-axis
(y-axis) is in unit of $r_{200}$ of the $X$ halo
($V_{200}$ for solid curves and $M_{200}$ for dashed curves).}
\end{figure}

Now let us examine the production of UHECR in the scenario
proposed here and the applicability of the virial theorem for
obtaining the $Y$ halo profile in Eq.~(\ref{yhalo}). It was found
that the level of the UHECR and the UHE neutrino fluxes produced
from $X$ decays is proportional to a single parameter $r_X=\xi_X
t_0/\tau_X$ for a fixed $m_X$, where $\xi_X$ is the present
fraction of $X$ particles in CDM and $t_0=13.7 {\rm Gyrs}$ is the
age of the Universe~\cite{ber} and there $r_X= 5\times
10^{-11}/\delta$ was used to fit the observed UHECR flux spectrum.
Note that a factor of $\delta$ is added because the energy of the
decay relativistic quarks and leptons is $E_{q,l}\sim\delta m_X$,
where $\delta\sim 1$ for the case in Ref.\cite{ber} and here
$\delta\sim 0.77v_X\sim 10^{-3}$ (where $v_X$ is about 300 km/s)
and $m_X= 10^{16} {\rm GeV}$, and also that the parameter $r_X$
will be larger if the energy dissipation of the decay particles is
taken into account~\cite{zia}. Assume that the $X$ halo is
originally formed at $0.1-1$ Gyrs and that $\tau_X=0.7 {\rm
Gyrs}$. Then the dynamical effect of $X$ decays on the halo is at
work from about $0.7$ Gyrs to the present time. Since $X$ and $Y$
are non-relativistic and $\tau_X<<t_0$, most $X$ particles have
decayed into $Y$ particles many Gyrs ago and the $Y$ DM halo has
been virialized. Otherwise, one should treat the recoil velocities
in a more proper way as considered in Ref.~\cite{sal} to estimate
the resulting halo profile. Hence we can see that we have found
out the allowed parameter space which is consistent with current
observation data and justified the method we used. In short, the
fraction of remaining $X$ particles in the recently formed $Y$ DM
halo is tiny and given by $\xi_X\sim 10^{-9}$, and they are
decaying at the present epoch to produce the observed UHECR
flux~\cite{ber}. Furthermore, the possible distortions of the
ionization history of the Universe caused by the energy injection
from decays of these relatively short-lived $X$ particles have
been recently discussed and the superheavy DCDM model is able to
provide a good fit to the current CMB anisotropy and polarization
data~\cite{doro}. On the other hand, the scenario proposed in
Ref.~\cite{cen}, where $\xi_X=1$ and $\tau_X \sim t_0$, would
produce unacceptably large flux of UHECR unless the relativistic
particles produced from the $X$ decay involve some exotic quarks
and leptons which are weakly interacting and may generate UHECR at
an acceptable level by interacting with the interstellar medium
when propagating to the Earth.

\section{Phenomenological implications }

We have shown that the KR model that has attempted to explain the presence
of UHECR with energies beyond the GZK cutoff can easily provide a
DCDM solution for the problem of subgalactic structure formation
in the CCDM model. In the DCDM model in which $X$ DM decay into
relativistic particles~\cite{cen}, not only halo core density is lowered
but also small dwarf galaxies are darkened due to core expansion and
subsequent quenched star formation. It has also been argued that presently
observed dwarf spheroidal galaxies with lower velocity dispersions were
resulted from decaying dark matter and subsequent core expansion in a
small fraction of halos with high velocity dispersions~\cite{cen2}.
This model predicts that the small-scale power at higher redshift
is enhanced compared to the CDM model as well as the gas fraction in clusters
should decrease with redshift. The latter can be tested by X-ray and
Sunyaev-Zel'dovich effect observations. However, this model has been criticized
for that the reduction in the central density of clusters of
galaxies due to $X$ DM evaporation might be too large to be compatible
with observations and could even be harmful to the halo
substructure formation~\cite{sal}. It has been pointed out
that this excessive reduction can be remedied if $X$ particles decay into
non-relativistic stable massive $Y$ DM, and shown that the $Y$ DM provides
well fits to the rotation curves of low-mass galaxies and
does not necessarily produce a significant reduction of the central DM
density of certain dwarf spheroidals~\cite{sal}. Undoubtedly, detailed
numerical simulations of the subgalactic structure formation in the
DCDM model versus high-quality observations on the properties of
subhalos and X-ray/Sunyaev-Zel'dovich effect of clusters would test
the DCDM model and should differentiate the two scenarios.
Remarkably, the subhalo astrophysics at kpc scales may provide a hint to
understand the mass difference between $X$ and $Y$ in the KR model at
energy scale of grand unification.

To test models of superheavy particles directly in terrestrial particle
accelerators is quite impossible. However, the particle spectra and
the arrival directions of UHECR produced from decays of superheavy particles
in the Galactic halo can provide crucial tests. Superheavy particles decay
into ultra high-energy quark and lepton jets which fragment predominantly
into photons with a small admixture of protons~\cite{kuz,ber}. Although UHECR
observations seem to show a subdominant photon flux~\cite{ave},
the photon flux with energies near the GZK cutoff may be attenuated in the
cascading of the jets in the radio background and intergalactic magnetic
fields~\cite{bha}. The ultra high-energy neutrino flux accompanying the
UHECR has been calculated~\cite{ber,bir,hoo} to be much
higher than the proton flux due to the long mean free path and high
multiplicity of neutrinos produced in high-energy hadronic jets. This
neutrino flux is near the detection limit of the on-going AMANDA neutrino
experiment and will be severely constrained by the upgraded AMANDA and
next generation neutrino telescope IceCube.
Because of the off-center location of the Solar system in the Galactic halo,
some amount of anisotropy in the arrival directions of UHECR
is expected~\cite{bha}. Recently it was claimed that no significant deviation
from isotropy is found, based on the data from the SUGAR and the AGASA
experiments taken a 10-year period with nearly uniform sky coverage~\cite{anch}.
This may be overturned due to insufficient statistics. It is likely that
the signal of the predicted anisotropy will have to wait to be tested by
the upcoming Pierre Auger Observatory.

As pointed out by KR~\cite{kuz}, instanton mediated decay processes
typically lead to multiparticle final states. Thus $X$ particle decays
will produce a relatively large number of quark jets with a fairly flat
energy distribution and rather hard leptons as compared to typical
perturbative decays of superheavy particles. This may leave a distinct
signature in the predicted UHECR spectrum which may help in distinguishing
the KR model from other DCDM models. Furthermore, in the KR model
which has $\delta \alt 1$, the energy of the relativistic $Y$ particle is
about $m_X/2$ and the flux of $Y$ particles in the Solar vicinity
is approximately given by $n_X R_{halo}/\tau_X\sim 10^{-5}n_X$ , where
$R_{halo}\sim 100 {\rm kpc}$ is the size of the Galactic halo. This
flux is about two orders of magnitude lower than the local flux of
typical halo DM which is estimated as
$n_X v_X\sim 10^{-3}n_X$ (where $v_X$ is about 300 km/s).
If the $Y$ particle interacts weakly with ordinary matter, it
may scatter with the target nucleus with mass $m_N$ in a cyrogenic
detector and deposit a huge amount of energy of order
$m_N (1-\delta)^{-2}$ in the detector. This deposit energy is much larger
than that of a typical halo DM particle which is about $m_N v_X^2$.
This may give a new thought to the direct detection of halo DM.
Unfortunately, since the local number density of $X$ is
$n_X\sim ({\rm GeV}/m_X) {\rm cm}^{-3}$ strongly suppressed by the mass of $X$
and $X$ is weakly interacting, the direct search for halo $X$
particles or the indirect search for high-energy neutrinos from decaying
$X$ particles captured in the Sun or the Earth in current experiments
are elusive~\cite{jun}. However, it is worth noting that the fluxes of
$X$-induced high-energy neutrinos from the Sun and the Earth are expected
to be similar, though they are relatively low, to those
considered in a different context of annihilation of strongly interacting
superheavy DM which are predominantly tau neutrinos with a flat energy
spectrum of events at about few TeV~\cite{far} and distinguishable
from the energy spectrum of high-energy neutrinos induced by neutralino
DM~\cite{gai}.

\section{Conclusions and Discussions}

In conclusion we have discussed the implication of the Kuzmin-Rubakov's
decaying superheavy dark matter model for generating cosmic rays with
energies beyond the Greisen-Zatsepin-Kuzmin cutoff to the subgalactic
structure formation of the Universe. The model involving a new $SU(2)_X$
gauge interaction and two left-handed $SU(2)_X$ fermionic
doublets ${\bf X}$ and ${\bf Y}$ can easily accommodate decaying dark
matter scenarios for solving the cuspy halo problem inherent in the
collisionless cold dark matter model. Intriguingly, the longevity
of $X$ particles due to instanton mediated decays may explain
the presence of a small cosmological constant as well.

The drawback is that we require the near-degeneracy of $X$ and $Y$ 
particle masses. However, this may have recourse to physics at the
relevant high energy scale.
In order to obtain the near mass degeneracy between $X$ and $Y$
particles, we assume that at high energies there is a symmetry, for example
an exchange symmetry between $X$ and $Y$, that
makes their masses equal. Small mass differences could be generated by
radiative corrections from symmetry breaking terms arising
via threshold corrections near grand unification scale or even from
stringy effects near Planck scale. For example, consider a term 
$\lambda_1{\cal L}_1$ which contains $X$ and other heavy fields.
The one-loop correction lifts $X$ mass by a factor of $\lambda_1^2/{16\pi^2}m_X$,
giving rise to $\delta\sim 10^{-2}$ for $\lambda_1\sim 1$. 
To get an even smaller $\delta$, we may use 
the idea of collective breaking of symmetries. 
Instead of using one single coupling to break the symmetry, 
we introduce another similar coupling $\lambda_2{\cal L}_2$ in such a way 
that each coupling by itself preserves sufficient amount of symmetry such
that the mass degeneracy between $X$ and $Y$ is exact at one-loop level.
It is only when the simultaneous presence of both symmetry breaking terms the
mass degeneracy will be lifted. Therefore the radiative corrections which 
lift the mass degeneracy of $X$ and $Y$ are necessarily proportional to both
$\lambda_1$ and $\lambda_2$. Hence this mass degeneracy splitting
effect occurs at two-loop level and is of order 
$\lambda_1^2/{16\pi^2} \; \lambda_2^2/{16\pi^2}$
which is sufficiently small even for $\lambda_1 \sim \lambda_2 \sim 1$.
An alternative mechanism for generating a small mass difference between
$X$ and $Y$ particles is closely related to the result of instanton 
effects considered here. The mass relation between $X$ and $Y$ may be slightly
modified by nonperturbative mass renormalization due to instanton-induced
counterterms, similar to instanton-generated quark masses considered in 
QCD physics~\cite{choi}.

It is quite interesting to link different astrophysical and cosmological
problems in a single particle model at grand unification scale.
Future observations of dark matter halos and ultra high-energy cosmic rays,
halo dark matter experimental search, and future CMB anisotropy and
polarization measurements will test the decaying dark
matter models and shed light on the mass degeneracy of $X$ and $Y$.

\section{Acknowledgments}

The work of K.W.N. (C.H.C.) was supported in part by the National Science
Council, Taiwan, ROC under the Grant NSC91-2112-M-001-026
(NSC91-2811-M-001-048).


\begin{references}

\bibitem{wang}
See, e.g., L. Wang, R. R. Caldwell, J. P. Ostriker, and P. J. Steinhardt,
Astrophys. J. {\bf 530}, 17 (2000).
\bibitem{wmap}
C. L. Bennett {\it et al.}, Astrophys. J. Suppl. {\bf 148}, 1 (2003).
\bibitem{nfw}
J. F. Navarro, C. S. Frenk, and S. D. M. White,
Astrophys. J. {\bf 462}, 563 (1996).
\bibitem{moore}
B. Moore, F. Governato, T. Quinn, J. Stadel, and G. Lake,
Astrophys. J. {\bf 499}, L5 (1998).
\bibitem{ccdm}
A. A. Klypin, A. V. Kravtsov, O. Valenzuela, and F. Prada,
Astrophys. J. {\bf 522}, 82 (1999);
B. Moore, S. Ghigna, F. Governato, G. Lake, T. Quinn, J. Stadel, and P. Tozzi,
Astrophys. J. {\bf 524}, L19 (1999).
\bibitem{dwarf}
B. Moore, Nature (London) {\bf 370}, 629 (1994); R. A. Flores and
J. A. Primack, Astrophys. J. {\bf 427}, L1 (1994); A. Burkert,
Astrophys. J. {\bf 447}, L25 (1995); W. J. G. De Blok and S. S.
McGaugh, Mon. Not. R. Astron. Soc. {\bf 290}, 533 (1997); A.
Borriello and P. Salucci, Mon. Not. R. Astron. Soc. {\bf 323}, 285
(2001).
\bibitem{galaxy}
V. P. Debattista and J. A. Sellwood, Astrophys. J. {\bf 493}, L5
(1998) ; J. F. Navarro and M. Steinmetz, Astrophys. J. {\bf 528},
607 (2000); P. Salucci and A. Burkert, Astrophys. J. {\bf 537}, L9
(2000).
\bibitem{cluster}
J. A. Tyson, G. P. Kochanski, and I. P. Dell'Antonio,
Astrophys. J. {\bf 498}, L107 (1998).
\bibitem{sper}
D. N. Spergel and P. J. Steinhardt, Phys. Rev. Lett. {\bf 84}, 3760 (2000).
\bibitem{kap}
M. Kaplinghat, L. Knox, and M. S. Turner, Phys. Rev. Lett. {\bf 85}, 3335 (2000).
\bibitem{cen}
R. Cen, Astrophys. J. {\bf 546}, L77 (2001).
\bibitem{sal}
F. J. S$\acute{\rm a}$nchez-Salcedo, Astrophys. J. {\bf 591}, L107 (2003).
\bibitem{gzk}
K. Greisen, Phys. Rev. Lett. {\bf 16}, 748 (1966);
G. T. Zatsepin and V. A. Kuzmin, Pis'ma Zh. $\acute{{\rm E}}$ksp. Teor. Fiz.
{\bf 4}, 114 (1966) [JETP Lett. {\bf 4}, 78 (1966)].
\bibitem{fly}
D. J. Bird {\it et al.}, Astrophys. J. {\bf 511}, 739 (1999).
\bibitem{agasa}
N. Hayashida {\it et al.}, Astrophys. J. {\bf 522}, 225 (1999);
M. Takeda {\it et al.}, Phys. Rev. Lett. {\bf 81}, 1163 (1998).
\bibitem{ellis}
J. Ellis, G. B. Gelmini, J. L. Lopez, D. V. Nanopoulos, and S. Sarkar,
Nucl. Phys. B {\bf 373}, 399 (1992).
\bibitem{bir}
M. Birkel and S. Sarkar, Astropart. Phys. {\bf 9}, 297 (1998).
\bibitem{ber}
V. Berezinsky, M. Kachelriess, and A. Vilenkin,
Phys. Rev. Lett. {\bf 79}, 4302 (1997).
\bibitem{kuz}
V. A. Kuzmin and V. A. Rubakov, Yad. Fiz. {\bf 61}, 1122 (1998)
[Phys. At. Nucl. {\bf 61}, 1028 (1998)].
\bibitem{fod}
Z. Foder and S. D. Katz, Phys. Rev. Lett. {\bf 86}, 3224 (2001).
\bibitem{bla}
P. Blasi, R. Dick, and E. W. Kolb, Astropart. Phys. {\bf 18}, 57 (2002).
\bibitem{bha}
See, e.g., P. Bhattacharjee and G. Sigl, Phys. Rept. {\bf 327}, 109 (2000).
\bibitem{jai}
P. Jaikumar and A. Mazumdar, Phys. Rev. Lett. {\bf 90}, 191301 (2003).
\bibitem{yok}
J. Yokoyama, Phys. Rev. Lett. {\bf 88}, 151302 (2002).
\bibitem{chu}
D. J. H. Chung, E. W. Kolb, and A. Riotto, Phys. Rev. D {\bf 60}, 063504 (1999).
\bibitem{enq}
K. Enqvist, K.-W. Ng, and K. A. Olive, Nucl. Phys. B {\bf 303}, 713 (1988);
D. J. H. Chung, E. W. Kolb, and A. Riotto, Phys. Rev. D {\bf 59}, 023501 (1999);
V. Kuzmin and I. Tkachev, Phys. Rev. D {\bf 59}, 123006 (1999).
\bibitem{baa}
J. Baacke, K. Heitmann, and C. P$\ddot{{\rm a}}$tzold, Phys. Rev.
D {\bf 58}, 125013 (1998); P. Greene and L. Kofman, Phys. Lett. B
{\bf 448}, 6 (1999); G. F. Giudice, M. Peloso, A. Riotto, and I.
Tkachev, J. High Energy Phys. {\bf 08}, 014 (1999); R. Allahverdi
and M. Drees, Phys. Rev. Lett. {\bf 89}, 091302 (2002).
\bibitem{zia}
H. Ziaeepour, Astropart. Phys. {\bf 16}, 101 (2001).
\bibitem{doro}
A. G. Doroshkevich and P. D. Naselsky, Phys. Rev. D {\bf 65}, 123517 (2002);
A. G. Doroshkevich {\it et al.}, Astrophys. J. {\bf 586}, 709 (2003);
R. Bean, A. Melchiorri, and J. Silk, Phys. Rev. D {\bf 68}, 083501 (2003).
\bibitem{cen2}
R. Cen, Astrophys. J. {\bf 549}, L195 (2001).
\bibitem{ave}
M. Ave {\it et al.}, Phys. Rev. Lett. {\bf 85}, 2244 (2000).
\bibitem{hoo}
D. Hooper and F. Halzen, hep-ph/0110201; and references therein.
\bibitem{anch}
L. A. Anchordoqui {\it et al.}, Phys. Rev. D {\bf 68}, 083004 (2003).
\bibitem{jun}
See, e.g., G. Jungman, M. Kamionkowski, and K. Griest,
Phys. Rep. {\bf 267}, 195 (1996).
\bibitem{far}
A. E. Farraggi, K. A. Olive, and M. Prospelov, Astropart. Phys.
{\bf 13}, 31 (2000);
I. F. M. Albuquerque, L. Hui, and E. W. Kolb, Phys. Rev. D {\bf 64},
083504 (2001).
\bibitem{gai}
T. K. Gaisser, G. Steigman, and S. Tilav, Phys. Rev. D {\bf 34},
2206 (1986);
J. S. Hagelin, K.-W. Ng, and K. A. Olive, Phys. Lett. B {\bf 180},
375 (1986);
K.-W. Ng, K. A. Olive, and M. Srednicki, Phys. Lett. B {\bf 188},
138 (1987).
\bibitem{choi}
K. Choi, C. W. Kim, and W. K. Sze, Phys. Rev. Lett. {\bf 61}, 794 (1988).
\end{references}
\end{document}